\documentclass[superscriptaddress, twocolumn, amsmath, amssymb, aps, prl,longbibliography]{revtex4-2}
\usepackage{amsmath, amssymb, amsfonts}
\usepackage{graphicx}
\usepackage{dcolumn}
\usepackage{bm}
\usepackage[colorlinks,linkcolor=blue,anchorcolor=blue,citecolor=blue]{hyperref}
\usepackage{color}

\begin{document}

\title{Synthetic SU(1,1) Gauge Control of Non-Hermitian Light}

\author{Wanyue Xiao}
\affiliation{Department of Physics, City University of Hong Kong, Tat Chee Avenue, Kowloon, Hong Kong, China}

\author{Ruo-Yang Zhang}
\affiliation{National Laboratory of Solid State Microstructures, School of Physics, Collaborative
Innovation Center of Advanced Microstructures, Nanjing University, Nanjing, China}

\author{Yuqiong Cheng}
\affiliation{Department of Physics, City University of Hong Kong, Tat Chee Avenue, Kowloon, Hong Kong, China}

\author{Shubo Wang}
\email{shubwang@cityu.edu.hk}
\affiliation{Department of Physics, City University of Hong Kong, Tat Chee Avenue, Kowloon, Hong Kong, China}

\begin{abstract}
\noindent Synthetic gauge fields enable nontrivial light control but remain largely restricted to compact groups such as U(1) and SU(2). Here, we study paraxial light propagation under a noncompact SU(1,1) gauge field realized in a non-Hermitian medium. Under this field, light evolves pseudo-unitarily and conserves pseudo-intensity despite material gain or loss. The gauge field drives Lorentzian precession of the optical pseudospin on a Bloch hyperboloid, producing rotations, boosts, and exceptional-point critical dynamics in an internal Minkowski space. A non-Abelian Lorentz force converts these internal dynamics into real-space oscillatory, hyperbolic, and cubic pseudo-intensity centroid trajectories, even in a uniform medium, while light of opposite pseudo-intensity can drift oppositely—a non-Hermitian spin-Hall-like effect. Our results connect non-Hermitian wave physics with noncompact gauge dynamics and reveal hidden conserved quantities in non-Hermitian light transport.
\end{abstract}

\maketitle

\textit{Introduction.---}Local gauge symmetry, from U(1) electromagnetism to non-Abelian Yang--Mills theory \cite{Yang1954}, underpins modern physics. Synthetic gauge fields have enabled wave and particle control across electronic \cite{Xiao2010,Fujita2011}, atomic \cite{Lin2009,Dalibard2011}, photonic \cite{Fang2012PRL,Fang2012NP,Fang2013PRL,Liu2015,Chen2019,Yang2024,Song2025,Lai2025}, and acoustic \cite{xiao2015synthetic,wang2018topological} systems but remain largely restricted to compact groups such as U(1) and SU(2) \cite{Xiao2010,Fujita2011,Lin2009,Dalibard2011,Fang2012PRL,Fang2012NP,Fang2013PRL,Liu2015,Chen2019,Yang2024,Song2025,Lai2025,xiao2015synthetic,wang2018topological}. Extending this paradigm to noncompact groups can bring qualitatively new gauge dynamics but requires a different framework.

SU(1,1), the minimal noncompact counterpart of SU(2), is central to quantum-optical phenomena such as two-mode squeezing and parametric amplification, where it acts unitarily on an infinite-dimensional Fock space \cite{Barut1971,Perelomov1972,Yurke1986}. However, SU(1,1) admits no nontrivial finite-dimensional unitary representations and thus cannot be realized as a gauge symmetry in finite-dimensional unitary optical systems. Instead, engineered non-Hermiticity \cite{Bender1998,Mostafazadeh2002,Bender2002} circumvents this limitation by enabling finite-dimensional pseudo-unitary representations of SU(1,1) that preserve an indefinite inner product \cite{Bargmann1947,Azizov1989,Mostafazadeh2004,Mostafazadeh2006}. This opens a route to synthetic SU(1,1) gauge control in non-Hermitian optical systems \cite{Ruter2010,zhang2018dynamically,Miri2019,Ozdemir2019,wang2019arbitrary}, bringing intrinsic hyperbolic geometry and Lorentzian character of SU(1,1) into finite-dimensional optical realizations.

\begin{figure}[tb]
\centering
\includegraphics[width=\linewidth]{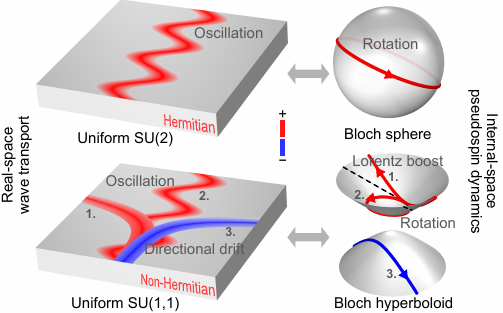}
\caption{Compact SU(2) versus noncompact SU(1,1) gauge dynamics in uniform media. Hermitian SU(2) gauge field drives compact pseudospin rotations on the Bloch sphere, corresponding to bounded oscillatory transport of conserved positive-definite intensity (top). Non-Hermitian SU(1,1) gauge field drives pseudospin rotations and noncompact Lorentz boosts on the Bloch hyperboloid, corresponding to oscillatory and directional transport of conserved pseudo-intensity (bottom). Red and blue denote positive and negative (pseudo-) intensity.}
\label{Fig1}
\end{figure}

In this Letter, we synthesize a real-space SU(1,1) gauge field for light in a non-Hermitian bianisotropic medium. In a uniform medium, SU(1,1) gauge symmetry renders paraxial light evolution pseudo-unitary, preserving electromagnetic pseudo-intensity while ordinary intensity grows or decays. With pseudo-intensity preserved, the gauge field drives the optical SU(1,1) pseudospin---the noncompact counterpart of SU(2) pseudospin (Bloch vector)---in an internal (2+1)D Minkowski space. The resulting dynamics are rotation-dominated, Lorentz-boost-dominated, and critical in unbroken, broken, and exceptional-point (EP) regimes, respectively, tracing closed elliptic, open hyperbolic, and parabolic trajectories on the Bloch hyperboloid. Through the non-Abelian Lorentz force, the evolving pseudospin steers the beam in real space, mapping internal Lorentzian dynamics onto oscillatory, hyperbolic, and cubic pseudo-intensity centroid trajectories. This real-sapce transport also exhibits a non-Hermitian spin-Hall-like effect. These noncompact internal and real-space dynamics under an SU(1,1) gauge field contrast with compact pseudospin precession and bounded beam oscillation under an SU(2) gauge field (Fig.~\ref{Fig1}). Our results reveal a conserved transport channel in non-Hermitian light that is controlled by the SU(1,1) gauge field, exhibits noncompact dynamics, and is invisible to conventional intensity measurements. 

\textit{Noncompact SU(1,1) gauge fields for light.---}We consider monochromatic electromagnetic waves in a \(z\)-invariant medium with constitutive relations \(\mathbf{D} = \tensor{\varepsilon}\mathbf{E} + \tensor{\chi}_{\mathrm{em}}\mathbf{H}\), \(\mathbf{B} = \tensor{\mu}\mathbf{H} + \tensor{\chi}_{\mathrm{me}}\mathbf{E}\), whose background response satisfies \(\tensor{\varepsilon}/\varepsilon_0 = \alpha\tensor{\mu}/\mu_0 = \text{diag}(\tensor{\varepsilon}_T,\varepsilon_z)\), \(\tensor{\chi}_{\mathrm{em}} = \tensor{\chi}_{\mathrm{me}} = 0\). Here \(\tensor{\varepsilon}_T\) and \(\varepsilon_z\) are real, and \(\alpha>0\). We set \(\alpha=1\); results for \(\alpha \neq 1\) follow by rescaling \(\varepsilon_0 \rightarrow \alpha\varepsilon_0\). Source-free Maxwell equations in this background possess a global SL(2,\(\mathbb{C}\)) symmetry \cite{SM}, extending the conventional SO(2) electromagnetic-duality rotation between
\(\mathbf{E}\) and \(\mathbf{H}\) \cite{Liu2015,Bliokh2013,FernandezCorbaton2012}. We introduce material-tensor modulations:
\begin{equation}
\begin{split}
\frac{\Delta\tensor{\varepsilon}}{\varepsilon_0} = \begin{pmatrix} 0 & i\mathbf{h}_1 \\ i\mathbf{h}_1^\dagger & 0 \end{pmatrix},\;\;
\frac{\Delta\tensor{\mu}}{\mu_0} = \begin{pmatrix} 0 & i\mathbf{h}_2 \\ i\mathbf{h}_2^\dagger & 0 \end{pmatrix}, \\
c\Delta\tensor{\chi}_{\mathrm{em}} = c\Delta\tensor{\chi}_{\mathrm{me}} = \begin{pmatrix} 0 & -\mathbf{x} \\ -\mathbf{x}^{\mathrm{T}} & 0 \end{pmatrix}.
\end{split}
\label{eq:1}
\end{equation}
The complex in-plane vectors $\mathbf{h}_{i}=(h_{ix},h_{iy})^{\mathrm{T}}~(i=1,2)$ specify anti-Hermitian off-diagonal electric and magnetic responses and satisfy \(\mathbf{h}_2=-\mathbf{h}_1^*\); the real vector $\mathbf{x}$ denotes lossless Tellegen magnetoelectric coupling. In this 2D medium, Maxwell equations can be recast into an in-plane wave equation
\begin{equation}
\hat{\mathcal{H}}|\psi\rangle = \left[\frac{1}{2}(\hat{\mathbf{p}}^{\mathrm{T}}-\hat{\mathcal{A}}^{\mathrm{T}})\tensor{\mathbf{m}}^{-1}(\hat{\mathbf{p}}-\hat{\mathcal{A}}) + V_0\right]|\psi\rangle = 0.
\label{eq:2}
\end{equation}
Here \(|\psi\rangle = (E_z,\eta_0 H_z)^{\mathrm{T}}\) is the SU(1,1) spinor, with \(\eta_0 = \sqrt{\mu_0/\varepsilon_0}\); \(\hat{\mathbf{p}} = -i(\partial_x,\partial_y)^{\mathrm{T}}\), \(V_0 = -k_0^2\varepsilon_z\), and \(\tensor{\mathbf{m}} = (\tensor{\varepsilon}_T^{\mathrm{T}})^{-1}\det(\tensor{\varepsilon}_T)/2\) are the canonical momentum operator, U(1) scalar potential, and effective anisotropic mass, respectively. The SU(1,1) gauge potential enters the effective Hamiltonian \(\hat{\mathcal{H}}\) of \(|\psi\rangle\) through minimal coupling \(\hat{\mathbf{p}} \rightarrow \hat{\mathbf{p}}-\hat{\mathcal{A}}\) and decomposes as \(\hat{\mathcal{A}}= \mathcal{A}^{a}\hat{\tau}_{a}\ = (\mathcal{A}_{x}^{a},\mathcal{A}_{y}^{a})^{\mathrm{T}}\hat{\tau}_{a}\) (\(a=1,2,3\)), where \(\mathcal{A}_x^a,\mathcal{A}_y^a \in \mathbb{R}\). Its components are related to material tensors by \(\mathcal A^{1}=k_0\operatorname{Re}(\mathbf h_1)\times\mathbf e_z\), \(\mathcal A^{2}=k_0\operatorname{Im}(\mathbf h_1)\times\mathbf e_z\), and \(\mathcal A^{3}=k_0\mathbf x\times\mathbf e_z\). The SU(1,1) generators \((\hat{\tau}_1,\hat{\tau}_2,\hat{\tau}_3) = (i\hat{\sigma}_1,i\hat{\sigma}_2,\hat{\sigma}_3)\) satisfy \([\hat{\tau}_b,\hat{\tau}_c] = 2if^{a}_{\;bc}\hat{\tau}_a\), with \(f^{3}_{\;12}=-1\), \(f^{2}_{\;31}=1\), \(f^{1}_{\;23}=1\), and \(f^{a}_{\;bc}=-f^{a}_{\;cb}\).
In 2D, the gauge field strength is the non-Abelian
magnetic field
\(
\hat{\mathcal{B}}
= \mathcal{B}^a\hat{\tau}_a
= \left(\nabla\times\mathcal{A}^a
+ f^{a}_{\;bc}\mathcal{A}^b\times\mathcal{A}^c\right)\hat{\tau}_a
\), with \(\mathcal{B}^a\)
normal to the $xy$-plane. Because the potential components do not commute, \(f^{a}_{\;bc}\mathcal{A}^b\times\mathcal{A}^c\)
can persist even for uniform $\mathcal{A}^a$. At each point, the two-component spinor $|\psi\rangle$ defines an SU(1,1) pseudospin vector,
\begin{equation}
s^a(\mathbf r)=
\frac{\langle\psi|\hat\tau_3\hat\tau^a|\psi\rangle}
{\langle\psi|\hat\tau_3|\psi\rangle},
\label{eq:5}
\end{equation}
with $\hat\tau^a=\eta^{ab}\hat\tau_b$ and $\eta^{ba}\eta_{ac}=\delta^b_{\;c}$. $s^a$ is a Lorentz vector in an internal $(2+1)$D Minkowski space spanned by generators $\{\hat{\tau}_a\}$ with metric $\eta_{ab}=\mathrm{diag}(-1,-1,1)$, where $\hat\tau_{1,2}$ and $\hat\tau_3$ are spacelike and timelike, respectively. The SU(1,1) pseudospin is the noncompact counterpart of the SU(2) Bloch vector and satisfies $s_a s^a=\eta_{ab}s^a s^b=1$  \cite{AD}, thus lying on a unit two-sheeted hyperboloid---the Bloch hyperboloid (Fig.~\ref{Fig1}).

In a gauge description, the local frame of $|\psi\rangle$ is nonunique; Eq.~(\ref{eq:2}) is covariant under a local SU(1,1) gauge transformation $|\psi'\rangle=\hat G(\mathbf r)|\psi\rangle$, with $\hat G^\dagger\hat\tau_3\hat G=\hat\tau_3$. Unlike SU(2), this transformation is pseudo-unitary and preserves the indefinite-metric inner product, $\langle\psi'|\hat\tau_3|\psi'\rangle=\langle\psi|\hat\tau_3|\psi\rangle$, defining the gauge-invariant local pseudo-intensity $\rho(\mathbf r)
=
\langle\psi|\hat\tau_3|\psi\rangle
=
|E_z|^2-|\eta_0H_z|^2.$ It can be positive or negative, whereas the ordinary local intensity $I(\mathbf r)=\langle\psi|\psi\rangle=|E_z|^2+|\eta_0H_z|^2$ is positive and gauge-dependent. The same gauge transformation induces a Lorentz transformation of $S^a=\langle\psi|\hat\tau_3\hat\tau^a|\psi\rangle$ $(s^a=S^a/\rho,\ \rho\neq0)$ in the internal Minkowski space, preserving the Minkowski length $S_aS^a=\rho^2$. The above describes the local SU(1,1) gauge transformation; for light viewed as a semiclassical particle carrying an internal spinor degree of freedom, propagation under the gauge field corresponds to continuous SU(1,1) evolution, so that its pseudo-intensity is dynamically conserved.

\begin{figure}[t!]
\centering
\includegraphics[width=\linewidth]{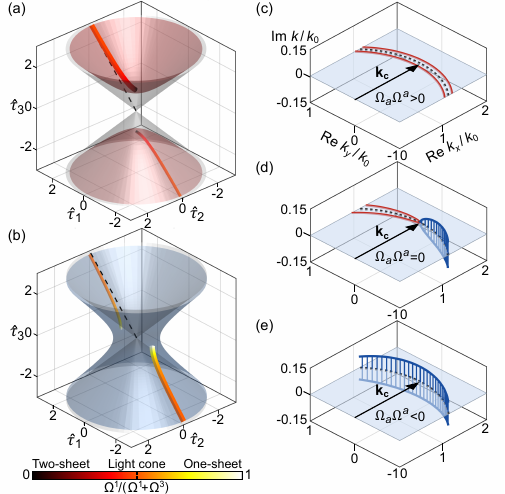}
\caption{(a,b) Two-/one-sheeted hyperboloids (red/blue) for timelike/spacelike \(\Omega^a\), and a light cone (gray) for lightlike \(\Omega^a\). Eigenstates are represented by biorthogonal projectors \(\hat{P}_\pm = \frac{1}{2}(\hat{\tau}_0\pm n^a\hat{\tau}_a)\) in (a) and \(\hat{P}_\pm = \frac{1}{2}(\hat{\tau}_0\mp in^a\hat{\tau}_a)\) in (b), corresponding to antipodal points \(\pm n^a\) on respective hyperboloids; thick/thin curves show
representative loci of \(\hat P_+/\hat P_-\) as \(\Omega^{1,3}\) vary. At the EP, projectors collapse to a single null direction on the light cone (black dashed line). (c--e) Complex isofrequency contours for \(\mathcal{A}^1/k_0 = (0.05,0)^\mathrm{T}\), \((0.1,0)^\mathrm{T}\), \((0.1,0)^\mathrm{T}\) and \(\mathcal{A}^3/k_0 = (0.1,0)^\mathrm{T}\), \((0.1,0.1)^\mathrm{T}\), \((0.05,0)^\mathrm{T}\), with selected \(\mathbf{k}_\mathrm{c}\) (black arrow) satisfying \(\Omega_a\Omega^a>0\), \(=0\), and \(<0\), respectively. Black dashed curve: unsplit contour \(h_0(\mathbf{k}_\mathrm{c})=0\).}
\label{Fig2}
\end{figure}

\begin{figure*}[t]
\centering
\includegraphics[width=\linewidth]{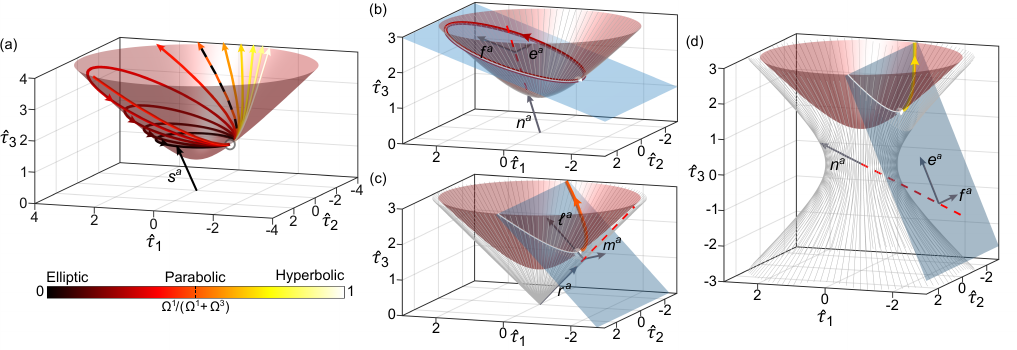}
\caption{(a) Pseudospin trajectories on the Bloch hyperboloid \(s_as^a=1\) as \(\Omega^{1,3}\) vary; dashed line: trajectory for \(\mathbf{k}_\mathrm{c}\) at the EP, white dot: initial pseudospin. (b,d) Unbroken/broken phases with timelike/spacelike \(n^a\) on wireframe unit two-/one-sheeted hyperboloids \((n_an^a=\pm1)\); blue planes \(\mathrm{span}\{e^a,f^a\}\) are Euclidean/(1+1)D Minkowski, and white pseudospin orbits are their intersections with the Bloch hyperboloid (red). Each plane passes through \(s^a_\parallel\) and is parallel to the tangent plane of the wireframe hyperboloid at \(n^a\). (c) At the EP, projectors collapse to the null direction \(\ell^a \propto \Omega^a\) on the light cone, and the blue plane is the shifted degenerate plane \(\ell^\perp = \text{span}\{\ell^a,m^a\}\), with a constant Jordan-companion offset along \(r^a\).}
\label{Fig3}
\end{figure*}

\textit{Internal Lorentzian causal structure of paraxial light.---}For an optical beam in a uniform SU(1,1) gauge field with isotropic mass \(\tensor{\mathbf{m}} = m\hat{\tau}_0\) (\(m \in \mathbb{R}, \hat{\tau}_0\) the identity), the paraxial approximation of Eq.~(\ref{eq:2}) yields the Schr\"odinger-type equation along propagation parameter \(\zeta\) \cite{SM}:
\begin{equation}
i\frac{\mathrm{d}}{\mathrm{d}\zeta}|\tilde{\psi}\rangle = \hat{H}(\mathbf{k}_\mathrm{c})|\tilde{\psi}\rangle,\;\hat{H}(\mathbf{k}_\mathrm{c})\!= \!h_0(\mathbf{k}_\mathrm{c})\hat{\tau}_0 + \frac{1}{2}\Omega^a(\mathbf{k}_\mathrm{c})\hat{\tau}_a,
\label{eq:4}
\end{equation}
with \(h_0(\mathbf{k}_\mathrm{c}) = (\mathbf{k}_\mathrm{c}\cdot\mathbf{k}_\mathrm{c}+C)/2m + V_0\), \(C = |\mathcal{A}^3|^2 - |\mathcal{A}^2|^2 - |\mathcal{A}^1|^2\) and \(\Omega^a(\mathbf{k}_\mathrm{c}) = -\frac{2}{m}\mathbf{k}_\mathrm{c}\cdot\mathcal{A}^a\). Here \(|\tilde{\psi}\rangle\) is the slowly varying envelope, \(|\psi\rangle = |\tilde{\psi}\rangle\exp(i\mathbf{k}_\mathrm{c}\cdot\mathbf{r})\), with $\mathbf{k}_\mathrm{c}$ the central wave vector. Under the SU(1,1) gauge field, the traceless part of \(\hat{H}\) is \(\mathfrak{su}(1,1)\)-valued, making \(\hat{H}\) pseudo-Hermitian \cite{Mostafazadeh2010,Grimaudo2018,Deng2025}, \(\hat{\tau}_3\hat{H}\hat{\tau}_3 = \hat{H}^\dagger\). Consequently, paraxial evolution is pseudo-unitary and preserves the pseudo-intensity \(\mathcal{N} =\int \mathrm{d}\mathbf{r}_{\perp} \langle\tilde{\psi}(\mathbf{r}_\perp)|\hat{\tau}_3|\tilde{\psi}(\mathbf{r}_\perp)\rangle=\int \mathrm{d}\mathbf{r}_\perp\,\rho(\mathbf{r}_\perp)\), even as the ordinary intensity \(\int \mathrm{d}\mathbf{r}_{\perp}\,\langle\tilde{\psi}(\mathbf{r}_\perp)|\tilde{\psi}(\mathbf{r}_\perp)\rangle=\int \mathrm{d}\mathbf{r}_\perp\,I(\mathbf{r}_\perp)\) grows or decays (integrals are over the transverse beam cross section).
\(\Omega^a\) is an effective Lorentz vector in the internal Minkowski space. Since \(\Omega^a = -\frac{2}{m}\mathbf{k}_\mathrm{c}\cdot\mathcal{A}^a\), the synthetic gauge field determines this Lorentz vector and its causal type, which is geometrically represented by three surfaces (Table~\ref{table}). For timelike \(\Omega^a\) (\(\Omega_a\Omega^a>0\)) and spacelike \(\Omega^a\) (\(\Omega_a\Omega^a<0\)), the normalized direction \(n^a = \Omega^a/\sqrt{|\Omega_b\Omega^b|} \in \mathbb{R}^3\) lies on a two-sheeted hyperboloid (\(n_an^a=1\)) in Fig.~\ref{Fig2}(a) and a one-sheeted hyperboloid (\(n_an^a=-1\)) in Fig.~\ref{Fig2}(b), respectively; for lightlike \(\Omega^a\), \(\Omega_a\Omega^a=0\) gives the light cone.

\begin{table}[b!]
\caption{\label{tab:causal}
Causal classification of $\Omega^a$.}
\setlength{\tabcolsep}{4pt}
\begin{ruledtabular}
\begin{tabular}{@{}cccc@{}}
$\Omega_a\Omega^a$ & Causal type & Causal surface & Phase \\
\hline
$>0$ & timelike  & two-sheeted hyperboloid & unbroken \\
$=0$ & lightlike (null) & light cone              & EP        \\
$<0$ & spacelike & one-sheeted hyperboloid  & broken
\end{tabular}
\end{ruledtabular} \label{table}
\end{table}

\(\hat{H}(\mathbf{k}_\mathrm{c})\) has two eigenstates whose eigenvalues \(\lambda_\pm(\mathbf{k}_\mathrm{c}) = h_0(\mathbf{k}_\mathrm{c}) \pm \frac{1}{2}\sqrt{\Omega_a\Omega^a}\) determine two eigen-wavevectors on the corresponding isofrequency contours. The two eigen-wavevectors are quasi-degenerate around the real central wave vector \(\mathbf{k}_\mathrm{c}=k_{\mathrm{c}}\hat{\mathbf{k}}_{\mathrm{c}}\). $\mathbf{k}_\mathrm{c}$ lies on the real contour \(h_0(\mathbf{k}_\mathrm{c})=0\) in complex $\mathbf{k}$-space, i.e., \(|\mathbf{k}_\mathrm{c}|=\sqrt{-2mV_0-C}\), a one-dimensional circle parametrized only by the direction \(\hat{\mathbf{k}}_{\mathrm{c}}\) [Fig.~\ref{Fig2}(c--e), dashed line]. Along \(\hat{\mathbf{k}}_{\mathrm{c}}\), the two eigen-wavevectors are separated by \(\delta k = (\lambda_+-\lambda_-)/v_g = \sqrt{\Omega_a\Omega^a}/v_g\), with \(v_g = k_{\mathrm{c}}/m\). The causal type of \(\Omega^a\) governs \(\delta k\): it is real for timelike \(\Omega^a\) [unbroken phase, Fig.~\ref{Fig2}(c)], imaginary for spacelike \(\Omega^a\) [broken phase, Fig.~\ref{Fig2}(e)], and zero at the EP for lightlike \(\Omega^a\), where the two isofrequency contours coalesce at $\mathbf{k}_{\mathrm{c}}$ [Fig.~\ref{Fig2}(d)]. This quasi-degenerate approximation requires \(|\delta k|\ll|k_{\mathrm{c}}|\). The eigenstates corresponding to the two eigen-wavevectors are superposed to form $|\tilde{\psi}\rangle$. They are represented in internal Minkowski space by biorthogonal projectors \(\hat{P}_\pm\) \cite{SM}, corresponding to antipodal points \(\pm n^a\) on the respective hyperboloid. At the EP, the two eigenvalues and eigenstates coalesce, and the projectors collapse to a single null direction $\ell^a\!\propto\!\Omega^a$ on the light cone.

\begin{figure*}[t!]
\centering
\includegraphics[width=\linewidth]{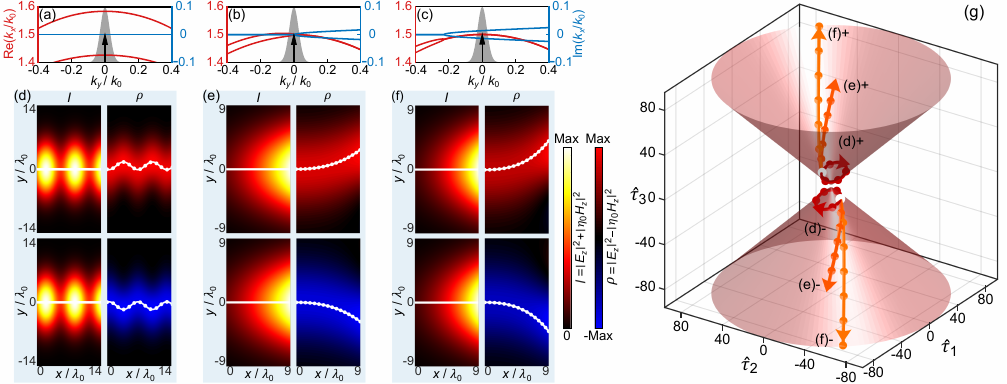}
\caption{(a--c) Isofrequency contours for \(\mathcal{A}^1/k_0 = (0.033,0.138)^\mathrm{T}\), \((0.022,0.022)^\mathrm{T}\), \((0.026,0.026)^\mathrm{T}\) and \(\mathcal{A}^3/k_0 = (0.085,0.053)^\mathrm{T}\), \((0.022,0)^\mathrm{T}\), \((0.022,0)^\mathrm{T}\), respectively. \(\text{diag}(\tensor{\varepsilon}_T,\varepsilon_z) = \text{diag}(1.5,1.5,1.5)\). Here \(k_y\) is real and \(k_x\) complex for the angular spectrum of a
finite-width beam propagating along \(x\), centered at \(k_y=0\).
Gray profile: incident angular spectrum
(beam width \(8\lambda_0\), \(\lambda_0=2\pi/k_0\)). (d--f) Beam evolution for initial spinors \((1.9,-1.7)^\mathrm{T}\) (upper) and \((-1.7,1.9)^\mathrm{T}\) (lower). \(\rho\) and \(I\) columns show simulated pseudo-intensity and intensity distributions. Solid curves show simulated
\(Y(x)=\int y\rho\,\mathrm{d}y/\int\rho\,\mathrm{d}y\) and intensity
centroids \(\int yI\,\mathrm{d}y/\int I\,\mathrm{d}y\) evaluated at fixed $x$; dots show
analytical \(Y(x)\). (g) Pseudospin trajectories for (d--f); upper/lower sheets: positive/negative pseudo-intensity; dots/curves: theory/simulations.
}
\label{Fig4}
\end{figure*}
\textit{Pseudospin dynamics in internal Minkowski space.---}The optical beam carries an SU(1,1) pseudospin $s^a=\langle\hat{\tau}^a\rangle_{1,1}$ on the Bloch hyperboloid (distinct from the causal hyperboloid in Fig.~\ref{Fig2}(a)). The gauge field drives its Lorentzian precession:
\begin{equation}
\frac{\mathrm{d}s^a}{\mathrm{d}\zeta} = i\langle[\hat{H},\hat{\tau}^a]\rangle_{1,1} = f^{a}_{\;bc}\Omega^b s^c = \mathsf{L}^{a}_{\;c}s^c.
\label{eq:6}
\end{equation}
Here \(\langle\hat{O}\rangle_{1,1} = {\int d\mathbf{r}_\perp\langle\tilde{\psi}(\mathbf{r}_\perp)|\hat{\tau}_3\hat{O}|\tilde{\psi}(\mathbf{r}_\perp)\rangle}/\mathcal{N}\) is the cross-section-averaged \(\hat{\tau}_3\)-metric expectation, with $\mathsf{L}^{a}_{\;c}=f^{a}_{\;bc}\Omega^b$ the (2+1)D Lorentz transformation generator. Equation~(\ref{eq:6}) identifies $\Omega^a$ as a Lorentzian generalization of the ``angular-velocity vector''. Unlike SU(2) pseudospin precession on the Bloch sphere, which contains only a compact Euclidean rotation, the SU(1,1) pseudospin precession on the Bloch hyperboloid  includes both the compact rotation generated by \(\Omega^3\) and the noncompact Lorentz boost generated by \(\Omega^{1,2}\). Accordingly, timelike, spacelike, and lightlike \(\Omega^a\) correspond to rotation-dominated, boost-dominated, and critical dynamics, with closed elliptic, open hyperbolic, and parabolic trajectories, as illustrated in Fig.~\ref{Fig3}(a) for \(\mathcal{N}>0\). For \(\mathcal{N}<0\), similar dynamics lie on the lower sheet and are not shown here.

Away from the EP, \(s^a\) decomposes into a longitudinal component \(s^a_\parallel\) along the precession axis \(n^a\) and a \(\zeta\)-dependent transverse component \(s^a_\perp\) in the plane spanned by \(e^a\) and \(f^a\) \cite{SM}. The vectors \(\{n^a,e^a,f^a\}\) form a Lorentzian orthonormal frame. In the unbroken phase [Fig.~\ref{Fig3}(b)], \(n^a\) is timelike, and \(\{e^a,f^a\}\) span a Euclidean plane; \(s^a_\parallel\) is determined by the eigenstate-weight imbalance, while \(s^a_\perp\) rotates in the Euclidean plane at constant angular rate \(\sqrt{\Omega_a\Omega^a}\), driven by relative phase evolution of the eigenstates. Consequently, \(s^a\) traces a closed elliptic trajectory on the Bloch hyperboloid. In the broken phase [Fig.~\ref{Fig3}(d)], \(n^a\) is spacelike and \(\{e^a,f^a\}\) span a (1+1)D Minkowski plane; each eigenstate now has vanishing pseudo-intensity, so \(s^a_\parallel\) is determined by their coherence rather than weights, while \(s^a_\perp\) boosts in the Minkowski plane at constant boost rate \(\sqrt{-\Omega_a\Omega^a}\), driven by relative amplitude evolution of the eigenstates. Consequently, \(s^a\) traces an open hyperbolic trajectory. At the EP [Fig.~\ref{Fig3}(c)], the eigenstates coalesce and the projectors collapse to a null direction \(\ell^a \propto \Omega^a\) on the light cone; its Lorentzian-orthogonal complement is \(\ell^\perp = \text{span}\{\ell^a,m^a\}\), where \(m^a\) is spacelike (\(\ell_am^a=0\)) and $\ell^a$ itself belongs to $\ell^\perp$; the trajectory is a parabola in this plane, with a constant shift along the Jordan-companion vector \(r^a\) that lies outside \(\ell^\perp\) (\(\ell_ar^a\neq0\)).

\textit{Light transport in real space.---}Alongside internal Minkowski-space pseudospin dynamics, the gauge field drives real-space light transport with conserved pseudo-intensity, governed by  Heisenberg equations
\begin{gather}
\frac{\mathrm{d}}{\mathrm{d}\zeta}\langle\hat{\mathbf{p}}\rangle_{1,1} = i\langle[\hat{H},\hat{\mathbf{p}}]\rangle_{1,1} = 0. \label{eq:7}\\ 
\frac{\mathrm{d}}{\mathrm{d}\zeta}\langle\hat{\mathbf{r}}\rangle_{1,1}\! =\! \langle\hat{\mathbf{v}}\rangle_{1,1}\! = i\langle[\hat{H},\hat{\mathbf{r}}]\rangle_{1,1} \!=\!\frac{1}{m}(\mathbf{k}_\mathrm{c}-\mathcal{A}^a\eta_{ab}s^b). \label{eq:8}
\end{gather}
In a uniform gauge field, \(\langle\hat{\mathbf p}\rangle_{1,1}\!=\!\mathbf k_\mathrm c\) is conserved. Yet kinetic velocity changes via pseudospin--gauge-field coupling. Equations ~(\ref{eq:6})--(\ref{eq:8}) yield a Newton-type equation
\begin{equation}
m\frac{\mathrm{d}^2}{\mathrm{d}\zeta^2}\langle\hat{\mathbf{r}}\rangle_{1,1} = \eta_{ab}\langle\hat{\mathbf{j}}^b\rangle_{1,1}\times\mathcal{B}^a = \frac{1}{m}\eta_{ab}s^b\mathbf{k}_\mathrm{c}\times\mathcal{B}^a,
\label{eq:9}
\end{equation}
where \(\hat{\mathbf{j}}^a = \frac{1}{2}\{\hat{\mathbf{v}},\hat{\tau}^a\} = \frac{1}{m}\hat{\mathbf{p}}\hat{\tau}^a - \frac{1}{m}\mathcal{A}^a\hat{\tau}_0\) is the pseudospin-current operator. Equation~(\ref{eq:9}) is therefore the non-Abelian generalization of the Lorentz force: gauge magnetic field \(\mathcal{B}^a\) acts on pseudospin current, not scalar-charge current. Equations~(\ref{eq:6}) and~(\ref{eq:9}) optically realize Wong's equations for a classical Yang--Mills charge \cite{Wong1970}, with pseudospin \(s^a\) as the SU(1,1) non-Abelian charge: the gauge field drives $s^a$, and $s^a$ steers the beam through the force, transferring internal pseudospin dynamics to real-space beam motion. Integrating Eq.~(\ref{eq:8}) or~(\ref{eq:9}) gives the pseudo-intensity
centroid trajectory:
\begin{equation}
\langle\hat{\mathbf{r}}\rangle_{1,1} = \frac{\int \mathrm{d}\mathbf{r}_\perp\,\mathbf{r}\rho(\mathbf{r})}{\int \mathrm{d}\mathbf{r}_\perp\,\rho(\mathbf{r})} = \frac{\mathbf{k}_\mathrm{c}}{m}\zeta - \frac{\mathcal{A}^a\eta_{ab}}{m}\int_0^\zeta s^b(\zeta')\mathrm{d}\zeta',
\label{eq:10}
\end{equation}
with \(s^a(\zeta) = [e^{\mathsf{L}\zeta}]^{a}_{\;c}s_0^c\) and \(s_0^c\) the initial pseudospin; closed-form expressions for \(\langle\hat{\mathbf{r}}\rangle_{1,1}\) are in Ref. \cite{SM}. The second term of Eq.~(\ref{eq:10}) couples pseudospin evolution to \(\langle\hat{\mathbf{r}}\rangle_{1,1}\), so that \(\langle\hat{\mathbf{r}}\rangle_{1,1}\) inherits the dynamical character of pseudospin in Minkowski space: for \(\Omega_a\Omega^a>0\), it exhibits bounded periodic oscillations; for \(\Omega_a\Omega^a<0\), it grows hyperbolically; and at the EP, \(\Omega_a\Omega^a=0\), it becomes polynomial in \(\zeta\), with terms up to cubic order.

To illustrate this, we consider gauge fields with \(\mathcal{A}^{1,3}\). Isofrequency contours are
shown in Figs.~\ref{Fig4}(a--c). The beam is incident along \(x\), with
\(\mathbf{k}_\mathrm{c}=k_0(1.5,0)^\mathrm{T}\). Along \(\mathbf{k}_\mathrm{c}\) (black arrow), Figs.~\ref{Fig4}(a--c) correspond to timelike,
lightlike, and spacelike regimes. Eliminating \(\zeta\) from
\(\langle\hat{\mathbf{r}}\rangle_{1,1}=[X(\zeta),Y(\zeta)]\)
gives the analytical centroid trajectory \(Y(X)\), in agreement with full-wave Berreman simulations \cite{Berreman1972} in Figs.~\ref{Fig4}(d--f) ($\rho$ columns), which correspond to (a--c)\cite{SM}. In each case, the upper/lower rows correspond to states of positive/negative pseudo-intensity; their pseudospins lie on opposite sheets of the Bloch hyperboloid and trace inversion-symmetric trajectories about the origin [Fig.~\ref{Fig4}(g)], thereby yielding mirror-symmetric pseudo-intensity centroid trajectories about the \(x\)-axis, while ordinary intensity centroids in the \(I\) columns remain at \(y=0\). This gives rise to a non-Hermitian spin-Hall-like effect.
Importantly, the pseudo-intensity,
\(\int \mathrm{d}\mathbf{r}_\perp\,\rho\), remains conserved under paraxial pseudo-unitary evolution. In contrast, the intensity
\(\int \mathrm{d}\mathbf{r}_\perp\,I\) is not conserved, as the medium
is non-Hermitian. Thus, the non-Abelian Lorentz force accompanying Lorentzian pseudospin dynamics curves real-space light transport, to which ordinary intensity measurements are blind.

\textit{Conclusion.---}We establish a framework for noncompact SU(1,1) gauge control of light in non-Hermitian bianisotropic media. Light's internal degree of freedom manifests as an SU(1,1) pseudospin evolving on a Bloch hyperboloid under a Lorentzian angular-velocity vector determined by the gauge field. The causal type of this vector governs elliptic, hyperbolic, or parabolic pseudospin dynamics and, through the non-Abelian Lorentz force, corresponding oscillatory, hyperbolic, or cubic motion of the pseudo-intensity centroid, revealing a hidden conserved transport channel in non-Hermitian light. The minimal model with \(\mathcal{A}^{1,3}\) can be implemented through anisotropic gain/loss \cite{Yuan2009,Jiang2011,Ye2014} and Tellegen magnetoelectric coupling \cite{Jacobs2015,Radi2016,Yang2025}, respectively, particularly on microwave or circuit platforms where these responses can be independently tuned. Beyond uniform fields, spatially varying SU(1,1) gauge fields may enable geometric phases \cite{berry1984quantal,cohen2019geometric,Xiao2026acoustic,cheng2025riemann} and Aharonov--Bohm-type effects \cite{Fang2012PRL,Chen2019,Yang2024}. Our framework suggests extensions to higher noncompact groups and other classical and quantum wave platforms, opening routes to engineering relativistic pseudospin dynamics, controlling wave transport, and exploring noncompact topological and geometric phenomena in non-Hermitian wave physics \cite{Deng2025,Zhang2019,Nasari2023,Hu2024,Dong2025}.

\textit{Acknowledgments.---}
The work described in this paper was supported by grants from the National Natural Science Foundation of China (No. 12322416) and the Research Grants Council of the Hong Kong Special Administrative Region, China (Project No. AoE/P-502/20).

\textit{Data availability}---The data that support the findings of this study are available from the corresponding author upon reasonable request.

\bibliography{apssamp}

\end{document}